\documentclass[%
reprint,
amsmath,amssymb,
aps,
pra,
onecolumn,
superscriptaddress
]{revtex4-2}

\usepackage[english]{babel}
\usepackage{graphicx}%
\usepackage{svg}
\usepackage{bbm}%
\usepackage{hyperref}%
\usepackage{cleveref}
\usepackage{xcolor}
\usepackage{physics}
\usepackage{relsize}
\usepackage{amsthm}
\usepackage{csquotes}
\usepackage{algorithm}
\usepackage{algorithmic}
\usepackage{todonotes}

\usepackage{ragged2e}

\usepackage{mathtools}

\usepackage{xr}
\crefname{claim}{Claim}{Claims}
\crefname{cor}{Corollary}{Corollaries}

\hypersetup{
  colorlinks=true,
  linkcolor=blue!55!black,
  citecolor=blue!55!black,
  urlcolor=blue!55!black
}

\begin{document}

\title{Loss-Resilient Quantum Networking with Dicke Entanglement}

\author{Bowen Wang}
\affiliation{\mbox{State Key Laboratory of Precision Spectroscopy, East China Normal University, Shanghai 200241, China}}

\author{Hongyu Yang}
\affiliation{\mbox{State Key Laboratory of Precision Spectroscopy, East China Normal University, Shanghai 200241, China}}

\author{Jiacheng Sun}
\affiliation{\mbox{State Key Laboratory of Precision Spectroscopy, East China Normal University, Shanghai 200241, China}}

\author{Wanping Zhou}
\affiliation{\mbox{State Key Laboratory of Precision Spectroscopy, East China Normal University, Shanghai 200241, China}}

\author{Jizhou Wu}
\affiliation{Quantum Science Center of Guangdong-Hong Kong-Macao Greater Bay Area (Guangdong), Shenzhen 518045, China}


\author{Yunlong Xiao}
\email{mathxiao123@gmail.com}
\affiliation{Institute of Advanced Intelligence and Computing (IAIC), Agency for Science, Technology and Research (A*STAR), 1 Fusionopolis Way, \#16-16 Connexis, Singapore 138632, Republic of Singapore}

\author{Dian Wu}
\email{dwu@lps.ecnu.edu.cn}
\affiliation{\mbox{State Key Laboratory of Precision Spectroscopy, East China Normal University, Shanghai 200241, China}}
\affiliation{Shanghai Branch, Hefei National Laboratory, Shanghai 201315, China}

\author{Jian Wu}
\email{jwu@phy.ecnu.edu.cn}
\affiliation{\mbox{State Key Laboratory of Precision Spectroscopy, East China Normal University, Shanghai 200241, China}}
\affiliation{Chongqing Key Laboratory of Precision Optics, Chongqing Institute of East China Normal University, Chongqing 401121, China}
\affiliation{\mbox{Collaborative Innovation Center of Extreme Optics, Shanxi University, Taiyuan, Shanxi 030006, China}}





\begin{abstract}

Multipartite entanglement can endow a quantum network with capabilities unavailable classically, yet their evolution under progressively reduced physical and informational access remains largely unexplored.
Here we realize the first eight-photon half-excitation Dicke state, with a fidelity of $0.922(11)$, and use it to implement two distinct network tasks:
open-destination teleportation and quantum telecloning.
An unknown state can be teleported from one agent to any of seven remote agents with fidelities of $0.926(7)-0.941(6)$, while seven approximate clones are distributed with a fidelity of $0.713(3)$, close to the optimal value of $5/7$.
The protocol remains above the classical limit under imperfect photon detection, transmission loss and missing information from participating agents, and outperforms the corresponding GHZ benchmarks.
As fewer agents contribute, the teleportation fidelity falls through a stepwise hierarchy whose two extremes recover open-destination teleportation and optimal telecloning, connecting both tasks within a single operational framework.
These results establish Dicke entanglement as a promising architecture for loss-resilient quantum networks, opening a route towards quantum technologies that retain useful functionality under realistic operating conditions.

\end{abstract}

\maketitle


Distributing quantum entanglement across distant systems is central to the development of quantum technologies, enabling communication~\cite{kimble2008quantum,wehner2018quantum}, computation~\cite{Cirac1999,Main2025} and sensing~\cite{Kmr2014,Guo2019} capabilities beyond those attainable with classical networks.
Moving from isolated links to a genuine network, however, requires quantum correlations to be established and controlled collectively across many spatially separated agents, making multipartite entanglement a fundamental resource for distributed functionality~\cite{pan2012multiphoton}.
Photons are particularly well suited to this task, providing mobile carriers of entanglement that can be distributed and measured across remote agents~\cite{flamini2019photonic,simon2017global,wang2020integrated}.
Rapid experimental progress is bringing this prospect closer to reality~\cite{chen2021integratednetwork,herreraValencia2026network,huang2026fusion}:
quantum teleportation~\cite{bennett1993teleporting} has advanced from laboratory demonstrations~\cite{bouwmeester1997experimental} to inter-island~\cite{ma2012teleportation}, metropolitan~\cite{valivarthi2016metropolitan} and satellite-scale links~\cite{ren2017satellite}, while multipartite photonic entanglement has been realized across progressively larger systems~\cite{lu2007graph,wang2016tenphoton}.

A loss-resilient quantum network must retain its function across every stage of operation. 
In photonic implementations, the multipartite resource can be degraded during state preparation when multiple photons enter the same detection path; 
photons can be lost during transmission; 
and measurement outcomes from assisting agents may be unavailable, leaving the receiver with incomplete information to determine whether teleportation should be accepted.
Although quantum error correction has been extensively developed to suppress transmission noise~\cite{grassl1997erasure,ralph2005loss,varnava2006loss,azuma2015repeaters,lu2008coding,bell2014graph,zhang2022shor}, far less is known about network operation when imperfections in photon detection affect the resource generation or when only partial information from assisting agents is available. 
This raises a fundamental question: 
can a quantum network continue to outperform classical strategies when these limitations arise at different stages of its operation?

\begin{figure}[t]
    \centering
    \includegraphics[width=\linewidth]{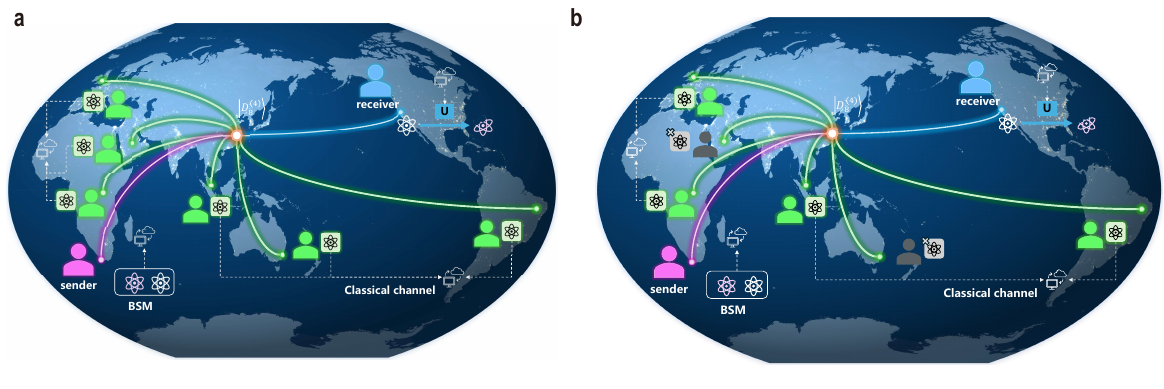}
    \caption{\protect\justifying
    \textbf{Open-Destination Teleportation.}
    \textbf{a,} Quantum network communication with the eight-photon Dicke state $\lvert D_8^{(4)}\rangle$. 
    The sender (magenta) performs a Bell-state measurement (BSM) between the input and its share of the Dicke state. 
    One of the seven remote agents is selected as the receiver (blue), while the other six act as assisting agents (green), whose local measurement outcomes are communicated through classical channels (dashed lines) and determine whether the teleportation event is accepted. 
    For accepted events, the BSM outcome specifies the local correction $U$ at the receiver.
    \textbf{b,} Operation with two assisting agents unavailable (grey). 
    Their measurement outcomes are absent, so acceptance is determined solely from the outcomes of the remaining agents. 
    Teleportation remains possible, with the missing assisting outcomes leading to a reduced fidelity.
    }
    \label{fig:theory}
\end{figure}

The answer lies in how multipartite entanglement is structured~\cite{dur2000three,briegel2001persistent,bourennane2006persistency,brunner2012persistency,barnea2015losses,neven2018robustness,glc7-xy8t}. 
Here, we extend the Dicke entanglement from previous four- and six-photon realizations~\cite{kiesel2007dicke,Schmid2008dick,prevedel2009dicke,wieczorek2009dicke,
schwemmer2014tomography,wu2026dicke,chen2023onchip} to an eight-photon state (Fig.~\ref{fig:theory}), and use it to demonstrate open-destination teleportation (ODT)~\cite{zhao2004odt,prevedel2009dicke,chiuri2012networking} and $1\!\rightarrow\!7$ telecloning~\cite{murao1999telecloning,prevedel2009dicke,chiuri2012networking}.
We test the network against incomplete information from assisting agents, imperfect photon detection during state preparation and photon loss during distribution.
Across these conditions, ODT remains above the classical limit~\cite{PhysRevLett.72.797,massar1995estimation} and exceeds the corresponding GHZ benchmarks.
As assisting-agent information is progressively removed, the fidelity does not decay smoothly but follows a stepwise hierarchy, with successive pairs of unavailable agents yielding the same performance. 
This hierarchy connects two seemingly distinct tasks: ODT when all assisting agents contribute, and optimal telecloning~\cite{gisin1997cloning,Werner1998} when none do.
Observing the same structure in six-photon Dicke states shows that the hierarchy is a property of Dicke entanglement rather than of a particular system size. 
Scalable quantum networking may therefore depend not simply on creating larger entangled states, but on identifying multipartite structures that continue to support network function when photons are lost or agents become unavailable.

\section*{RESULTS}


\subsection*{Dicke Entanglement for Quantum Communication}

\begin{figure}[t]
    \centering
    \includegraphics[width=\linewidth]{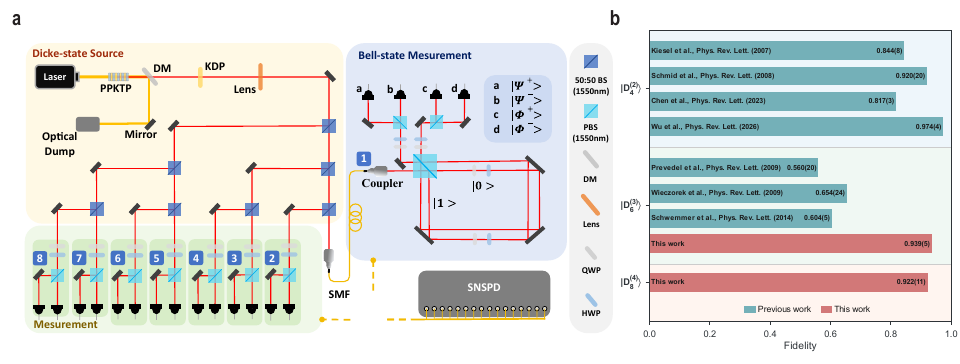}
    \caption{\protect\justifying
    \textbf{Experimental Setup and Benchmarking of Dicke States.}
    \textbf{a}, Experimental setup for generating the eight-photon Dicke state $\lvert D_8^{(4)}\rangle$ and performing the polarization-path Bell-state measurement. 
    A mode-locked Ti:sapphire laser (\(775~\mathrm{nm}\),
    \(80~\mathrm{MHz}\)) pumps a ppKTP crystal to produce four photon
    pairs via spontaneous parametric down-conversion. 
    A cascaded beam-splitter network distributes the photons into eight spatial modes. 
    Mode~1 is used for the Bell-state measurement (BSM) with the input qubit, while the other seven modes form the remote agents of the network.
    \textbf{b}, Fidelity benchmark for photonic Dicke-state generation. 
    Reported four- and six-photon results are shown alongside the six- and eight-photon states realized here. 
    The eight-photon Dicke state reaches a fidelity of 0.922(11), while the six-photon state reaches 0.939(5).
    }
    \label{fig:setup_fidelity}
\end{figure}

Realizing a quantum network in which a single sender can coherently connect to remote agents requires multipartite entanglement at both high fidelity and sufficient scale.
Here, we demonstrate the first experimental realization of an eight-photon Dicke state $\lvert D_{2n}^{(n)}\rangle$, generated through fourth-order emission from a collinear type-II spontaneous parametric down-conversion (SPDC) source (Fig.~\ref{fig:setup_fidelity} and Methods). 
This resource enables open-destination teleportation and quantum telecloning within a common architecture, establishing large-scale Dicke state as a versatile resource for quantum network communication.

In the experiment, one photon is assigned to the sender and the remaining seven polarization qubits are distributed among remote agents. 
Postselecting eightfold-coincidence events from the four-pair emission ideally projects the system onto the Dicke state
\begin{equation}
\lvert D_8^{(4)}\rangle
=
\frac{1}{\sqrt{70}}
\sum_{j=1}^{70}
P_j\bigl(
\lvert H\rangle^{\otimes4}
\otimes
\lvert V\rangle^{\otimes4}
\bigr),
\label{rev:eq_D8}
\end{equation}
where $P_j$ runs over all $\tbinom{8}{4}=70$ distinct permutations of polarized photons.
Eightfold coincidences are recorded at a rate of approximately $3.5\times10^{-2}\,\mathrm{s}^{-1}$.
We characterize the resulting state using an exact decomposition of the Dicke state projector into 29 collective polarization settings~\cite{Toth2009}, obtaining
\begin{equation}
F=0.922(11),
\label{rev:eq_FD}
\end{equation}
where the statistical uncertainty denotes one standard deviation. 
This fidelity lies well above the biseparable bound of $4/7$, corresponding to an entanglement witness value of $4/7-F=-0.351(11)$ and certifying genuine eight-partite entanglement.
The decomposition relies only on the symmetry of the target projector and makes no permutation-symmetry assumption about the prepared state; 
the effects of threshold detection and higher-order emission are treated separately in the Supplementary Information (SI).

\begin{figure}[b]
    \centering
    \includegraphics[width=\linewidth]{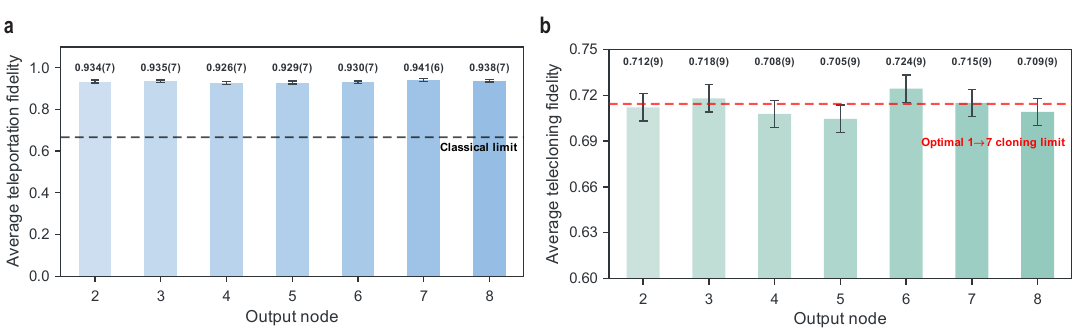}
    \caption{\protect\justifying
    \textbf{Teleportation and Telecloning Fidelities.}
    \textbf{a}, Open-destination teleportation fidelity for each of the seven possible receivers, evaluated over the uniformly sampled input ensemble $\mathcal{E}=\{\ket{H}, \ket{V}, \ket{D}, \ket{A}, \ket{R}, \ket{L}\}$.
    All receiver choices remain well above the classical limit of 2/3 (black dashed line).
    \textbf{b}, Average $1\!\rightarrow\!7$ quantum telecloning at each of the seven remote agents. 
    The red dashed line marks the optimal universal cloning fidelity of 5/7.
    Error bars denote one standard deviation from counting statistics.
    }
    \label{fig:odt_telecloning_node_fidelities}
\end{figure}

With the eight-photon Dicke state in hand, we turn to open-destination teleportation (Fig.~\ref{fig:theory}a), in which the receiver is chosen only after the entangled state has been distributed.
Any one of the seven remote agents can be designated as the receiver, while the other six perform local measurements whose outcomes determine whether the teleported state is accepted~\cite{zhao2004odt,chiuri2012networking}.
Using the uniformly sampled input ensemble $\mathcal{E}\coloneqq\{\ket{H}, \ket{V}, \ket{D}, \ket{A}, \ket{R}, \ket{L}\}$ with $\ket{D/A}\coloneqq(\ket{H}\pm\ket{V})/\sqrt{2}$ and $\ket{R/L}\coloneqq(\ket{H}\pm i\ket{V})/\sqrt{2}$, we obtain average fidelities ranging from $0.926(7)$ to $0.941(6)$ across all seven receiver choices (Fig.~\ref{fig:odt_telecloning_node_fidelities}a), far above the classical limit of 2/3~\cite{PhysRevLett.72.797,massar1995estimation}.
This demonstrates high fidelity teleportation with a receiver that remains undetermined at the time of entanglement distribution, providing a flexible primitive for dynamically reconfigurable quantum networks.
We then use the Dicke state for a complementary task, $1\!\rightarrow\!7$ quantum telecloning~\cite{murao1999telecloning,chiuri2012networking}, in which approximate copies of an unknown input are delivered simultaneously to all seven agents. 
For the same input ensemble $\mathcal{E}$, the measured average fidelity (Fig.~\ref{fig:odt_telecloning_node_fidelities}b),
\begin{equation}
\overline F_{\mathrm{cl}}=0.713(3),
\label{rev:eq_Fcl}
\end{equation}
is consistent with the optimal theoretical value of $5/7$~\cite{BuzekHillery1996,gisin1997cloning,Werner1998}.
For the average fidelity considered here, uniformly sampling the six Pauli eigenstates is equivalent to the Haar average over all pure states on the Bloch sphere~\cite{Bowdrey2002}. 
Details of the telecloning protocol are provided in Methods. 
Together, these results show that eight-photon Dicke entanglement can both route quantum information to a chosen destination and share it across the network.


\subsection*{Network Function under Information Loss}

A central challenge for quantum networks is maintaining their function when information from some agents becomes unavailable. 
In open-destination teleportation, the receiver relies on measurement outcomes from the other agents to decide whether the teleported state should be accepted. 
For multipartite GHZ entanglement, losing access to even a single subsystem reduces the teleportation fidelity to the classical limit of 2/3, eliminating the quantum advantage. 
We therefore ask how Dicke entanglement responds as information from progressively more agents becomes unavailable. 
This reveals a hierarchy of teleportation fidelities and, more fundamentally, a direct connection between open-destination teleportation and quantum telecloning.

\begin{figure}[htbp]
   \centering
   \includegraphics[width=\linewidth]{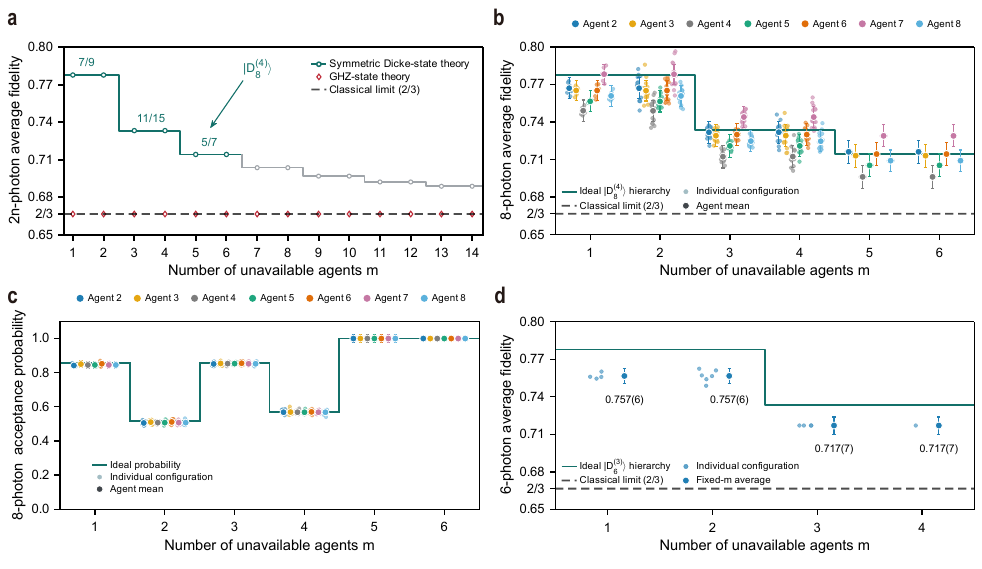}
   \caption{\protect\justifying
   \textbf{Network Performance Hierarchy under Agent Unavailability.}
   \textbf{a}, Theoretical teleportation fidelity as a function of the number $m$ of unavailable agents. 
   For Dicke state $\lvert D_8^{(4)}\rangle$, its fidelity follows a stepwise hierarchy;
   the three levels are 7/9, 11/15 and 5/7.
   The GHZ benchmark drops to the classical limit of 2/3 once any single agent becomes unavailable.
   \textbf{b}, Experimental realization of the hierarchy with the eight-photon Dicke state. 
   Points of different colors show the individual fidelities for the seven receivers, while dark points show the corresponding averages at fixed receiver and fixed $m$; 
   the solid line marks the ideal Dicke state values.
   \textbf{c}, Post-selection acceptance probabilities for the same eight-photon experiment. 
   Neighboring odd-even values of $m$ yield the same fidelity but generally different probabilities of acceptance.
   \textbf{d}, Independent test with the six-photon Dicke state $\lvert D_6^{(3)}\rangle$. 
   Dashed lines in \textbf{a, b} and \textbf{d} indicate the classical limit of 2/3.
   Error bars denote one standard deviation.
   }
   \label{fig:node_unavailability_fidelity}
\end{figure}

Our analysis relies on conservation of the total excitation number in the $2n$-photon Dicke state.
Suppose that, besides the sender and receiver, measurement outcomes from $m$ agents are unavailable. 
If these $m+2$ systems contain $\ell$ excitations, they occupy the Dicke state $\lvert D_{m+2}^{(\ell)}\rangle$, which contains $\tbinom{m+2}{\ell}$ equally weighted terms. 
For the sender and receiver to share the Bell state $\lvert\Psi^+\rangle\coloneqq(\ket{01}+\ket{10})/\sqrt{2}$, exactly one excitation must lie in this pair. 
There are two choices for its location, while the remaining $\ell-1$ excitations can be distributed among the $m$ unavailable agents in $\tbinom{m}{\ell-1}$ ways. 
The corresponding weight is therefore
\begin{equation}
p_{\Psi^+}(m,\ell)
=
\frac{2\binom{m}{\ell-1}}{\binom{m+2}{\ell}}
=
\frac{2\ell(m+2-\ell)}{(m+1)(m+2)}.
\label{eq:pPsi}
\end{equation}
For a given $m$, this probability is maximal when $\ell$ is closest to $(m+2)/2$, yielding $p_{\max}(m)=(m+2)/2(m+1)$ for even $m$ and $p_{\max}(m)=(m+3)/2(m+2)$ for odd $m$.
Writing $m=2r$ and $m=2r-1$, respectively, both cases give the same value, $p_{\max}(r)=(r+1)/(2r+1)$.
At this optimum, the teleportation fidelity depends on the input polarization:
for $\ket{H}$ and $\ket{V}$, $F_H(m)=F_V(m)=p_{\max}(m)$, whereas for $\ket{D}, \ket{A}, \ket{R}, \ket{L}$, it is $F_D(m)=F_A(m)=F_R(m)=F_L(m)=(1+p_{\max}(m))/2$.
Averaging over the ensemble $\mathcal{E}$ then gives $\overline F(m)=(1+2p_{\max}(m))/3$.
Consequently, each neighboring odd-even pair shares the same average fidelity,
\begin{align}
    \overline F(2r-1)=
    \overline F(2r)=
    \frac{2}{3}+\frac{1}{3(2r+1)},
    \quad r=1,\ldots,n-1,
\label{eq:degradationLaw}
\end{align}
forming a stepwise hierarchy as information from more agents becomes unavailable (Fig.~\ref{fig:node_unavailability_fidelity}a).
For the eight-photon Dicke state ($2n=8$), it falls from unity to 7/9 when one or two agents are missing, to 11/15 when three or four are missing, and finally to 5/7 when five or six are missing.

When all agents are available, the protocol realizes open-destination teleportation. 
As agents become unavailable, the network continues to operate, with its fidelity following the stepwise hierarchy derived above. 
At the opposite extreme, all $2n-2$ assisting agents are unavailable, so the receiver must rely entirely on its own share of the Dicke state. 
Equation~\eqref{eq:degradationLaw} then gives
\begin{align}
    \overline F(2n-1)=
    \frac{2(2n-1)+1}{3(2n-1)},
\label{eq:FendM}
\end{align}
which coincides with the fidelity of optimal universal $1\!\rightarrow\!(2n-1)$ quantum telecloning~\cite{prevedel2009dicke,gisin1997cloning,Werner1998}. 
Viewed across all $2n-1$ possible receivers, each remote share therefore carries an optimal approximate copy of the input state. Open-destination teleportation and quantum telecloning thus emerge as the two ends of the same framework: the former when all assisting agents can participate, and the latter when none of them can. The intermediate cases connect these limits through the stepwise loss of network performance.

To test the fidelity hierarchy predicted by Eq.~\eqref{eq:degradationLaw} across the entire network, we fix agent 1 as the sender and, for each receiver, vary the availability of the other six agents.
These six agents define $2^6$ possible availability patterns; 
excluding the all available situation leaves $2^6-1=63$ cases with at least one unavailable agent. 
Repeating this analysis for each of the seven possible receivers yields $7\times63=441$ fidelity estimates in total.
Each fidelity is reconstructed from the measured coincidence data, with statistical uncertainties obtained from $10^4$ Poisson Monte Carlo samplings.
In line with Eq.~\eqref{eq:degradationLaw}, the performance exhibits a clear three-step hierarchy:
$0.749$--$0.778$, $0.712$--$0.744$ and $0.696$--$0.729$ for one or two, three or four, and five or six unavailable agents (Fig.~\ref{fig:node_unavailability_fidelity}b), respectively, with uncertainties of $0.008$--$0.009$ (Methods).
Equal fidelity, however, does not imply operational equivalence:
although $2r-1$ and $2r$ unavailable agents give the same fidelity, they generally lead to different acceptance probabilities at the receiver.
For $m=0, \ldots, 6$, the corresponding probabilities are $(4/7, 6/7, 18/35, 6/7, 4/7, 1, 1)$, in agreement with the experimental data (Fig.~\ref{fig:node_unavailability_fidelity}c).
We further test the same behavior with a six-photon Dicke state, with fidelity $0.939(5)$ and a sixfold coincidence rate of $0.67\,\mathrm{s}^{-1}$.
The measured fidelities are $0.757(6)$ for $m=1,2$
and $0.717(7)$ for $m=3,4$, only $0.021$ and $0.016$ below the theoretical values
(Fig.~\ref{fig:node_unavailability_fidelity}d).
Observing the same stepwise structure in both the six- and eight-photon systems shows that the hierarchy in network performance is not tied to a particular system size, but is a general feature of Dicke entanglement as agents become unavailable.


\subsection*{Imperfect Generation and Lossy Distribution}

\begin{figure}[t]
    \centering
    \includegraphics[width=\linewidth]{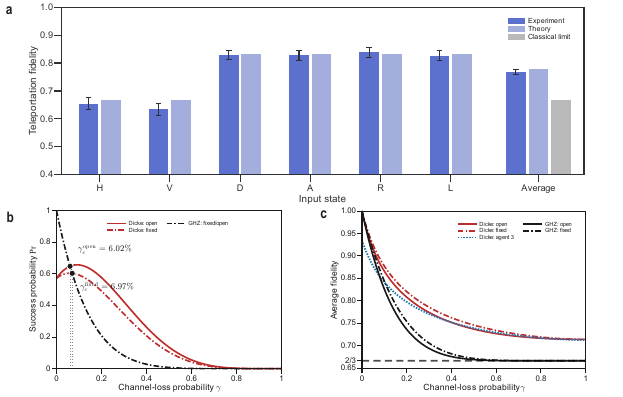}
    \caption{\protect\justifying
    \textbf{Network Performance under Failed Detection and Lossy Distribution.}
    \textbf{a}, Open-destination teleportation with a failed detection during Dicke state preparation. 
    The path to agent 8 is blocked and agent 3 is chosen as the receiver. 
    Dark blue bars show the experimental data for the input ensemble $\mathcal{E}$, and the light blue bars show the corresponding theoretical values. 
    The measured average fidelity is 0.768(8), above the classical limit of 2/3 (grey).
    \textbf{b}, Probability of successful communication as a function of the independent loss probability $\gamma$, taking a fidelity of 3/4 as the success threshold.
    Red curves show open- and fixed-destination teleportation with the Dicke state, while the black curve gives the corresponding GHZ benchmark. 
    The marked intersections define the loss probabilities beyond which the Dicke resource provides the higher success probability.
    \textbf{c}, Average fidelity after distribution through lossy channels of strength $\gamma$.
    Red curves denote the Dicke state for open- and fixed-destination teleportation, the blue dotted curve to agent 3 as the designated receiver, and black curves the corresponding GHZ benchmarks. 
    The grey dashed line marks the classical limit of 2/3.
    }
    \label{fig:physical-network-loss}
\end{figure}

Beyond the loss of agent information considered above, photonic networks face imperfections at two distinct stages: 
resource generation and subsequent distribution.
At the generation stage, the eight-photon Dicke state is obtained through coincidence postselection, so the resulting state is directly affected by detection failures.
Such failures can arise when an optical path is blocked or when higher-order emission sends multiple photons into the same path.
After preparation, a separate source of degradation arises when photons are lost during transmission through noisy quantum channels. 
We first examine the generation stage by blocking the path to agent 8 and choosing agent 3 as the receiver. 
Despite the missing detection event, the average teleportation fidelity remains $0.768(8)$ (Fig.~\ref{fig:physical-network-loss}a), above the classical limit of 2/3. 
A source-and-detection model gives a sevenfold-to-eightfold coincidence-rate ratio of 4.81, close to the measured value of 4.77;
further details are provided in SI.

Once prepared, the multipartite state must be distributed to the remote agents, where photon loss becomes the dominant source of degradation. 
We model each transmission channel by an independent loss probability $\gamma$ and compare fixed-destination with open-destination teleportation. 
To quantify the resilience of the network, we set a fidelity threshold of 3/4 and evaluate the probability that communication remains above this level. 
For this threshold, only events with $m\leq2$ unavailable agents contribute.
The corresponding success probabilities are
\begin{align}
    \mathrm{Pr}_{\text{fixed}}(\gamma)
    &=
    (1-\gamma)\sum_{m=0}^{2}
    \binom{6}{m}
    \gamma^m(1-\gamma)^{6-m}p_m,\\
    \mathrm{Pr}_{\text{open}}(\gamma)
    &=
    \sum_{m=0}^{2}
    \binom{7}{m}
    \gamma^m(1-\gamma)^{7-m}p_m,
\end{align}
where $m$ denotes the number of unavailable agents and $p_m$ gives the probability that the remaining measurement outcomes meet the acceptance condition.
For a GHZ resource, by contrast, quantum advantage is retained only when all seven transmitted photons survive, giving $\mathrm{Pr}_{\text{GHZ}}(\gamma)=(1-\gamma)^7$ for both protocols.
At $\gamma=0.2$, the Dicke state gives success probabilities of $0.491$ and $0.576$ for fixed- and open-destination teleportation, respectively, compared with $0.210$ for GHZ (Fig.~\ref{fig:physical-network-loss}b).

After transmission through channels with loss probability $\gamma=0.2$, we use the distributed Dicke and GHZ states for fixed- and open-destination teleportation. 
The average fidelities are 0.821 and 0.808 for the Dicke state, compared with 0.754 and 0.737 for GHZ (Fig.~\ref{fig:physical-network-loss}c). 
Fixed-destination teleportation performs slightly better because only the channel to the selected receiver contributes directly to the loss, whereas open-destination teleportation depends on successful distribution to all possible receivers. 
Across both tasks, the Dicke state maintains the higher fidelity, showing that its multipartite entanglement structure better preserves network performance under photon loss.


\section*{Discussions}

The first experimental realization of an eight-photon Dicke state advances both the scale of multipartite entanglement and the quantum networks it can support. 
Within a single architecture, this resource enables open-destination teleportation and $1\!\rightarrow\!7$ telecloning, allowing quantum information to be directed to a chosen receiver and shared across the network. 
Its function persists even when participating agents become inaccessible: 
the teleportation fidelity falls through a stepwise hierarchy whose two extremes recover open-destination teleportation and telecloning as limiting cases of a common framework. 
The same hierarchy emerges in six-photon Dicke states, indicating that it originates from the structure of Dicke entanglement itself. 
Useful quantum communication also survives imperfect detection during resource generation and photon loss during distribution, with Dicke entanglement outperforming the corresponding GHZ benchmarks in both settings.

Scaling Dicke entanglement to ten photons and beyond would allow quantum information to be directed to a larger pool of receivers, with the destination chosen after entanglement distribution according to task requirements and agent availability. 
Larger resources could also support quantum secret sharing protocols involving more participants, extending the role of multipartite entanglement from teleportation to the cooperative recovery of protected information.
Realizing these capabilities will require advances in photon generation, collection and detection, together with protocols that preserve useful performance under loss and incomplete local information. 
A central question is therefore how entanglement should be structured to meet the demands of a given network task. 
Extending the present framework to other multipartite resources and measurement strategies could help identify structures that sustain these functions under realistic operating conditions, guiding the development of larger quantum networks with useful and reliable capabilities.

\section*{Methods}

\subsection*{Eight-photon Dicke-state Generation and Characterization}

The eight-photon resource is generated from the four-pair component of a collinear type-II spontaneous parametric down-conversion (SPDC) source. A mode-locked Ti:sapphire laser delivers $140~\mathrm{fs}$ pulses at a central wavelength of $775~\mathrm{nm}$ and a repetition rate of $80~\mathrm{MHz}$. The laser pumps a periodically poled KTiOPO$_4$ (PPKTP) crystal, producing degenerate photon pairs near $1550~\mathrm{nm}$ with orthogonal horizontal ($H$) and vertical ($V$) polarizations. At the operating pump power, the single-pair generation probability is approximately $0.039$ per pulse.

Before spatial splitting, we compensate polarization-dependent temporal walk-off and optimize the spatial and spectral overlap of the two polarization components. A cascaded network of non-polarizing beam splitters distributes the four-pair component among eight spatial modes. Within the four-pair component, events with one photon in each mode constitute the intended eight-photon Dicke sample. No narrowband spectral filters are used in the output channels. Superconducting nanowire single-photon detectors have efficiencies of approximately $90\%$ at $1550~\mathrm{nm}$. With an $8~\mathrm{ns}$ coincidence window, the mean collection-and-detection efficiency across the eight modes is approximately $0.72$, and the observed eightfold coincidence rate is approximately $3.5\times10^{-2}\,\mathrm{s}^{-1}$.

Because the eight-photon measurements require extended acquisition, the polarization-domain two-photon interference and long-term stability of the source are monitored independently. At the operating temporal overlap, interference between the orthogonally polarized photons suppresses coincidences between the two orthogonal analyser outputs in the $\hat X$ basis. For each spatial mode $i$, we measure coincidence counts $N_Z^{(i)}$ and $N_X^{(i)}$ in the $\hat Z$ and $\hat X$ bases under otherwise identical conditions and equal integration times. We define the contrast as $R_{ZX}^{(i)}=N_Z^{(i)}/N_X^{(i)}$. For distinguishable photons, the expected $\hat X$-basis coincidence count is $N_Z^{(i)}/2$, giving $R_{ZX}^{(i)}=2$. The corresponding visibility is $V_{\mathrm{2ph}}^{(i)}=1-2/R_{ZX}^{(i)}$. The arithmetic means of the eight separately calculated contrasts and visibilities remain above $150$ and $0.987$, respectively. These measurements use a fixed temporal overlap rather than a delay-scanned Hong--Ou--Mandel dip. During acquisition, the pump power remains between $2.95$ and $3.10~\mathrm{W}$, while the collection-and-detection efficiencies of the eight modes show small long-term variations. Further stability data are provided in the Supplementary Information.

To characterize the resource, we evaluate the Dicke-state projector $\lvert D_8^{(4)}\rangle\langle D_8^{(4)}\rvert$ using an exact local decomposition into collective polarization observables, following Ref.~\cite{Toth2009}. We measure 29 collective polarization settings over an accumulated integration time of $174~\mathrm{h}$, obtaining approximately 750 eightfold events per setting on average. In each setting, the same polarization basis is applied to all eight modes, and the spatially resolved polarization outcomes are recorded. Measurements in the $H/V$ basis determine the excitation-number populations, while complementary bases provide the coherence contributions needed to evaluate the projector. The reduced number of settings exploits the symmetry of the target projector without assuming that the prepared state itself is permutation invariant. The decomposition and count analysis are provided in the Supplementary Information.

\subsection*{Six-photon Dicke-state Resource}

An independently prepared six-photon Dicke state, $\lvert D_6^{(3)}\rangle$, tests the fidelity hierarchy at a second system size. Its target-state fidelity is $0.939(5)$, evaluated using the same projector-decomposition approach~\cite{Toth2009}, and its observed sixfold coincidence rate is $0.67~\mathrm{s}^{-1}$. The source and experimental setup are described in the Supplementary Information.

For the communication measurements, one resource photon remains with the sender, one remote agent is selected as the receiver, and the other four agents provide assisting measurements. We apply the same protocol and information-loss analysis as in the eight-photon experiment. For the selected receiver, the four assisting agents define $2^4-1=15$ configurations with at least one unavailable measurement outcome. The excitation-number acceptance condition for each configuration is specified by the protocol before the measured fidelity is evaluated.

\subsection*{Input-state Preparation and Bell-state Measurement}

At the sender, a Sagnac interferometer maps the polarization qubit of one Dicke-resource photon onto two paths. Wave plates in the two paths then prepare the same input polarization state. Path thus carries the sender's resource qubit, while polarization carries the input qubit, allowing a Bell-state measurement on two degrees of freedom of one photon. Neither the six- nor the eight-photon experiment requires an additional input photon. The measured path-interference contrast is approximately $300{:}1$.

We prepare the six input states $\mathcal{E}=\{\lvert H\rangle,\lvert V\rangle,\lvert D\rangle,\lvert A\rangle,\lvert R\rangle,\lvert L\rangle\}$, where $\lvert D/A\rangle=(\lvert H\rangle\pm\lvert V\rangle)/\sqrt{2}$ and $\lvert R/L\rangle=(\lvert H\rangle\pm i\lvert V\rangle)/\sqrt{2}$. The polarization--path Bell-state analyser distinguishes the four Bell outcomes. The main communication measurements select the $\lvert\Psi^+\rangle$ outcome, which requires no receiver correction. In a separate verification with agent~3 as the receiver, we select the $\lvert\Psi^-\rangle$ outcome and apply the required Pauli-$Z$ correction before measuring the receiver.

For each input $\lvert\phi\rangle$, we measure the receiver in the basis $\{\lvert\phi\rangle,\lvert\phi^\perp\rangle\}$. The transfer fidelity is calculated from the accepted coincidence counts as
\begin{equation}
F_\phi=\frac{N_\phi}{N_\phi+N_{\phi^\perp}},
\end{equation}
where $N_\phi$ and $N_{\phi^\perp}$ are the counts for the target and orthogonal outcomes, respectively. The six input states are acquired separately, and the reported six-state fidelity is the equal-weight mean of their input-state fidelities.

In the ideal half-excitation Dicke model, the prescribed assisting-measurement condition is accepted with probability $\alpha_m$ when $m$ assisting-agent outcomes are unavailable. This probability is independent of the input state. Conditioned on assisting acceptance, the sender's resource qubit remains maximally mixed, so each Bell outcome occurs with probability $1/4$. The selected $\lvert\Psi^+\rangle$ branch therefore occurs jointly with assisting acceptance with probability $\alpha_m/4$ for every input.

Since this joint probability is input independent, normalizing the selected branch yields a trace-preserving qubit channel in the ideal model. Its equal-weight six-state fidelity equals the uniform Bloch-sphere average and can be compared with the $2/3$ measure-and-prepare benchmark. The derivation is given in the Supplementary Information. Experimental fidelities are calculated directly from accepted counts, without a correction based on the ideal branch probability.

\subsection*{Analysis of Unavailable Assisting-agent Outcomes}

We acquire a separate polarization-resolved eightfold data set for each of the seven receivers and six input states, giving 42 independently acquired data sets. An unavailable assisting outcome is represented by summing over that agent's two polarization results in the recorded counts. For one assisting agent $j$, the sum uses
\begin{equation}
\lvert H\rangle_j\langle H\rvert_j
+
\lvert V\rangle_j\langle V\rvert_j
=
I_j.
\end{equation}
For a subset $S$ of $m$ assisting agents, all $2^m$ polarization assignments are summed at the count level before normalization. Within the eightfold-coincidence sample, this gives the retained-outcome statistics of the polarization state after tracing over $S$.

After marginalization, we group the remaining assisting outcomes by excitation number. The protocol specifies the central sector, or two symmetry-related central sectors, to be accepted for each configuration; this choice is fixed before the measured fidelities are evaluated. We calculate the conditional transfer fidelity from accepted counts. For each input and configuration, the measured acceptance fraction is the accepted count divided by the count in the selected Bell-state branch before the assisting condition is applied. The ideal acceptance probability $p_m$ is calculated separately from Dicke-state combinatorics. Neither acceptance quantity is included in the conditional fidelity.

For each receiver, the six assisting agents define $2^6-1=63$ configurations with at least one unavailable outcome, giving 441 configurations across seven receivers. We first calculate each configuration's fidelity as the equal-weight mean of the six separately acquired input-state fidelities. At a fixed receiver and a fixed number $m$ of unavailable outcomes, we then average the configuration fidelities with equal weight; counts from different configurations are not pooled. The corresponding analysis gives $2^4-1=15$ configurations for the selected receiver in the six-photon experiment.

These configurations use events already registered as eightfold coincidences. Marginalization discards polarization information from those records without changing which trials entered the eightfold sample. Blocking an assisting path before detection changes the coincidence sample and its event rate. We therefore analyse the sevenfold blocking data separately from the reconstructed eightfold configurations.

\subsection*{Telecloning Analysis}

For telecloning, the sender uses the same polarization--path Bell-state measurement as for teleportation, while all seven remote photons serve as clone outputs. We select the $\lvert\Psi^+\rangle$ outcome, for which the receiver correction is the identity. No assisting-measurement acceptance condition is imposed. Each agent's photon is analysed for the six input states using the target and orthogonal polarization measurements described above.

At agent $j$, the six-state mean cloning fidelity is
\begin{equation}
\overline{F}^{(j)}
=
\frac{1}{6}\sum_{\phi\in\mathcal{E}}
\frac{N_\phi^{(j)}}{N_\phi^{(j)}+N_{\phi^\perp}^{(j)}}.
\end{equation}
Here, $N_\phi^{(j)}$ and $N_{\phi^\perp}^{(j)}$ are the corresponding coincidence counts for the target and orthogonal outcomes. The reported $1\!\rightarrow\!7$ telecloning fidelity is the equal-weight mean of $\overline{F}^{(j)}$ over the seven agents.

The Dicke-state telecloning channel is input dependent, so its fidelity varies among input states. Its six-state mean can be compared with the optimal universal-cloning average, but agreement between these averages does not imply that the Dicke-state channel is universal. The channel and its uniform Bloch-sphere average are derived in the Supplementary Information.

\subsection*{Physical Single-mode Blocking}

We block the optical path to assisting agent~8 before detection and keep agent~3 as the receiver. Sevenfold coincidences are recorded from the remaining paths for each of the six input states. We select the $\lvert\Psi^+\rangle$ Bell-state branch and apply the $m=1$ acceptance condition to the five remaining assisting agents. For comparison, agent~8's polarization outcome is marginalized in the eightfold records, and the same Bell branch and acceptance condition are applied to both data sets.

Sevenfold coincidences do not identify a unique photon-occupation pattern. The four-pair component contributes events with one photon unavailable through ordinary loss or the blocked path. It can also produce a sevenfold event when the blocked path is empty and two photons of the same polarization occupy one remaining path. Threshold detectors cannot resolve this double occupation, while higher-order SPDC emission provides further contributions. We therefore calculate the blocking fidelities from all accepted sevenfold counts and analyse these records separately from the marginalized eightfold data.

A source-and-detection model includes SPDC emission order, photon routing, optical and detection loss, and threshold detection. It estimates the composition of the sevenfold sample and predicts the sevenfold-to-eightfold coincidence-rate ratio after the same $m=1$ acceptance condition is applied to both samples. No model-based subtraction is applied to the measured fidelities. The model and its comparison with the measured rates are given in the Supplementary Information.

\subsection*{Statistical Analysis}

Individual input-state fidelities are evaluated from two orthogonal outcomes, and their uncertainties follow binomial counting statistics. Because the six input states are acquired separately, we propagate their uncertainties to the six-state mean fidelity of each teleportation configuration. The same procedure is used for the six-state mean cloning fidelity at each agent and for fidelities measured in the blocking experiment.

Configurations reconstructed from the same polarization-resolved records share counts and are statistically correlated. To evaluate an average over configurations, we sample each independent underlying count once per Poisson Monte Carlo realization, using its measured value as the Poisson mean. All configurations and their fixed-$m$ average are then recalculated from the same simulated records. This procedure preserves the correlations between configurations.

For the overall telecloning fidelity, the seven output fidelities for each input state are derived from the same eightfold record. We resample that record once per realization and use it to recalculate all seven output fidelities before averaging. Resource-fidelity uncertainties are obtained from $10^4$ Poisson Monte Carlo realizations of the projector-decomposition measurements, with the fidelity recalculated in each realization.

State and process tomography use maximum-likelihood reconstruction. For each measurement setting, binomial bootstrap samples are generated with the measured total count held fixed, and the complete reconstruction is repeated for each sample. The standard deviation of the reconstructed values gives the statistical uncertainty. Further details are provided in the Supplementary Information. All reported uncertainties correspond to one standard deviation.

\section*{Data availability}

The data that support the findings of this study are available
from the corresponding authors upon reasonable request.

\section*{Acknowledgements}
This work was supported by Quantum Science and Technology-National Science and Technology Major Project (Grant No. 2025ZD0301000), by the National Natural Science Foundation of China under the General Program (Grant No. 12474487), Shanghai Pilot
Program for Basic Research (No. TQ20240204), the Science and Technology Commission of Shanghai
Municipality (No. 231c1402900), and Shanghai Science and Technology Innovation Action Plan (Grant No. 24LZ1400200).
Y.X. is supported by A*STAR under its Career Development Fund (C243512002).

\section*{Competing interests}

The authors declare no competing interests.

\section*{References}
\bibliography{references}

\end{document}